\documentclass[journal,twoside]{IEEEtran}

\usepackage{amsmath}
\usepackage{booktabs}
\usepackage[caption=false,font=footnotesize]{subfig}
\usepackage[TS1,T1]{fontenc}
\usepackage{graphicx}
\usepackage{grffile}
\usepackage{pgfplots}
\usepackage[detect-all]{siunitx}
\usepackage{tikz}
\usepackage{microtype}
\usepackage{hyperref}

\usetikzlibrary{arrows.meta,shapes.arrows,shapes.geometric,shapes.symbols,shapes.multipart}
\pgfplotsset{compat=newest}
\pgfplotsset{plot coordinates/math parser=false}
\pgfplotsset{tick label style={font=\footnotesize}, label style={font=\small}, legend style={font=\tiny}, title style={font=\small}}

\newcommand{\centring}{\centering}

\DeclareMathOperator{\vvec}{vec}

\begin{document}
\def\figureautorefname{Fig.}
\def\sectionautorefname{Section}
\def\subsectionautorefname{Section}
\def\subfigureautorefname{\figureautorefname}

\title{Compressive Sampling Using a Pushframe Camera}
\author{\thanks{This work was supported by the UK Space Agency under the CEOI-11 grant NSTP3-PF-031.  \textit{(Corresponding author: Stuart Bennett.)}}
Stuart Bennett, Yoann Noblet, Paul F. Griffin, Paul Murray, Stephen Marshall, John Jeffers, and Daniel Oi\thanks{S. Bennett, P. Murray and S. Marshall are with the Department of Electronic and Electrical Engineering, University of Strathclyde, Glasgow, G1 1XQ, UK (email: stuart.bennett.100@strath.ac.uk; paul.murray@strath.ac.uk; stephen.marshall@strath.ac.uk)}\thanks{Y. Noblet, P. F. Griffin, J. Jeffers and D. Oi are with the Department of Physics, University of Strathclyde, Glasgow, G4 0NG, UK (email: yoann.noblet@strath.ac.uk; paul.griffin@strath.ac.uk; john.jeffers@strath.ac.uk; daniel.oi@strath.ac.uk)}}

\markboth{}{Bennett \MakeLowercase{\textit{et al.}}: Compressive Sampling Using a Pushframe Camera}
\IEEEpubid{\parbox{.95\textwidth}{\phantom{foo}~\phantom{bar}\\\phantom{.}}}

\maketitle

\begin{abstract}
The recently described pushframe imager, a parallelized single pixel camera capturing with a pushbroom-like motion, is intrinsically suited to both remote-sensing and compressive sampling.  It optically applies a 2D mask to the imaged scene, before performing light integration along a single spatial axis, but previous work has not made use of the architecture's potential for taking measurements sparsely.  In this paper we develop a strongly performing static binarized noiselet compressive sampling mask design, tailored to pushframe hardware, allowing both a single exposure per motion time-step, and retention of 2D correlations in the scene.  Results from simulated and real-world captures are presented, with performance shown to be similar to that of immobile --- and hence inappropriate for satellite use --- whole-scene imagers.  A particular feature of our sampling approach is that the degree of compression can be varied without altering the pattern, and we demonstrate the utility of this for efficiently storing and transmitting multi-spectral images.
\end{abstract}


\section{Introduction}
\label{s:intro}

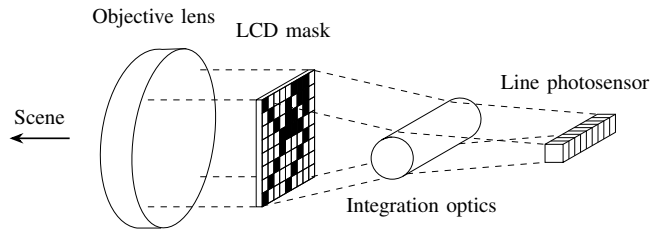
\begin{figure}
	\scalebox{.8}{
	\begin{tikzpicture}
		\draw[-Stealth, thick] (0.8,0) -- ++(-1.0, 0) node[midway,above=1mm] {Scene};

		\node[ellipse, draw, minimum height=3cm, minimum width=1.34cm, rotate=-8] (obj) at (2.53, 0) {};
		\draw (obj.south) -- ++(-.4, 0)[rotate=-8] arc[x radius = 8mm, y radius = 1.5cm, start angle = 270, end angle = 90] -- (obj.north);

		\draw (4,-1.139) -- (4,.639) -- (3.9,.639) -- (3.9,-1.139) -- (4,-1.139) -- (4.87,-.639) -- (4.87,1.139) -- (4.77,1.139) -- (3.9,.639);
		\draw (4,.639) -- (4.87,1.139);
		\def\pixelmap{{%
			{0,1,1,1,1,0,0,0,1},
			{1,0,1,1,0,1,0,0,1},
			{1,1,0,1,0,0,1,0,1},
			{1,1,1,0,0,0,0,1,1},
			{1,1,1,0,1,1,1,0,1},
			{1,1,0,1,1,1,0,1,1},
			{1,0,1,1,1,0,1,1,1},
			{0,1,1,1,0,1,1,1,1}}}
		\foreach \x in {0, ..., 8}
			\foreach \y in {0, ..., 7}{
				\pgfmathsetmacro{\bit}{\pixelmap[7-\y][\x]}
				\if\bit1
					\draw (4+.87/9*\x, -1.139+2/9*\y+.5/9*\x) -- (4.09666+.87/9*\x, -1.139+2/9*\y+.5/9+.5/9*\x) -- (4.09666+.87/9*\x, -.917+2/9*\y+.5/9+.5/9*\x) -- (4+.87/9*\x, -.917+2/9*\y+.5/9*\x) -- cycle;
				\else
					\draw[fill] (4+.87/9*\x, -1.139+2/9*\y+.5/9*\x) -- (4.09666+.87/9*\x, -1.139+2/9*\y+.5/9+.5/9*\x) -- (4.09666+.87/9*\x, -.917+2/9*\y+.5/9+.5/9*\x) -- (4+.87/9*\x, -.917+2/9*\y+.5/9*\x) -- cycle;
				\fi
			}

		\node[circle,draw,minimum size=7mm] at (6.16,-.33) {};
		\draw (6.16, .02) arc[radius = 3.5mm, start angle = 90, end angle = 120] -- ([turn]180:1.3) arc[radius = 3.5mm, start angle = 120, end angle = -60] -- ([turn]0:1.3);

		\draw (9,-.4) -- (9,-.1) -- (8.7,-.1) -- (8.7,-.4) -- (9,-.4) -- (9.87,.1) -- (9.87,.4) -- (9.57,.4) -- (8.7,-.1);
		\draw (9.87,.4) -- (9,-.1);
		\foreach \i in {1, ..., 8}
			\draw (8.7+.87/9*\i, -.1+.5/9*\i) -- (9+.87/9*\i, -.1+.5/9*\i) -- (9+.87/9*\i, -.4+.5/9*\i);

		\draw[dashed] (2.1,.639) -- (3.95,.639) -- (6.3,.1) -- (8.7,-.1);
		\draw[dashed] (2.1,-1.139) -- (3.95,-1.139) -- (6.05,-.66);
		\draw[dashed] (6.4,-.59) -- (8.7,-.4);
		\draw[dashed] (2.97,1.139) -- (4.82,1.139) -- (7.17,.6) -- (9.57,.4);
		\draw[dashed] (2.97,-.639) -- (3.9,-.639);
		\draw[dashed] (4.82,-.639) -- (5.82,-.41);
		\draw[dashed] (7.27,-.09) -- (8.95,.05);

		\node at (2.2,2) {Objective lens};
		\node at (4.35,1.8) {LCD mask};
		\node at (6.65,-1.15) {Integration optics};
		\node at (9.2,.9) {Line photosensor};
	\end{tikzpicture}
	}
	\caption{Basic pushframe optical path}
	\label{f:pfs}
\end{figure}

\IEEEPARstart{T}{he} `pushframe' camera concept, described by Noblet et al. in \cite{noblet2020}, may be viewed as a parallelized single pixel camera (SPC).  As in common SPC designs, a spatial light modulator (SLM) optically imposes a variable 2D mask on the incoming image, but rather than concentrating all unmasked light on a single photodiode, the pushframe architecture optically sums the light along one axis only, focusing each masked image column on to a separate photodiode, and so the image as a whole on to a 1D sensor array, as depicted in the \autoref{f:pfs} schematic.  If there is relative motion between the camera and the scene, similar to that required by a pushbroom sensor, a single appropriately designed 2D mask pattern can be constantly applied while a faithful image of the scene is stored, a coefficient being recorded for every scene column at every mask column.  Indeed, if the SLM pattern is a diagonal line, with the elements not forming the line being a blocking mask, the camera is equivalent to a pushbroom imager.  The cited application is use on a satellite platform for Earth observation, where a scanning pushbroom motion is easily achieved, and the rapid movement over the Earth makes a static pattern a near-necessity if any sort of competitive ground sampling distance is to be obtained.  The main benefit of the imager, relative to a pushbroom sensor, is improved signal to noise ratio, as multiple exposures may be combined to form the reconstructed image.

Noblet et al.'s architecture paper introduced a simple sampling scheme, based on a Walsh-Hadamard matrix, using a complete set of linearly independent one dimensional patterns, which achieved good reconstruction of the scene.  However, the described approach requires $n$ samples to reconstruct an $n$ pixel image: while there are SNR advantages, compared to the use of a pushbroom sensor, there would be no reduction in the data storage and transmission requirements.  The 2006 papers of Cand\`es et al. (\cite{candes2006}) and Donoho (\cite{donoho2006}) may be seen as establishing the field of compressive sampling (CS), where a signal is undersampled in a specific way, but then reconstructed near-perfectly through the use of prior knowledge of the signal's properties --- in particular that in some basis the signal has a sparse representation, and the samples have been taken in a different basis.  In the case of image capture, this takes the form of knowing that natural scenes have high compressibility --- i.e. sparseness --- in the wavelet domain.  Yuan and Haimi-Cohen's recent study (\cite{yuan2020}) compares image storage sizes, both compressed by CS (taking fewer samples) and the JPEG algorithm (quantized discrete cosine transform), and shows that, at 10:1 compression ratio, a CS reconstruction may have similar quality to that achieved by JPEG decompression, while both methods maintain 80--90\% similarity to the original image.  CS has been used in combination with SPCs (\cite{duarte2008}) to reduce the number of masks and exposures necessary to adequately reconstruct a scene, and hence reduce both capture time, and storage requirements.

\IEEEpubidadjcol
In this paper we describe a bespoke pushframe CS scheme.  The design of our algorithm is partially dictated by the combination of pushframe hardware and an Earth-observation application: for a given scene, the rapid scanning motion of the camera over the planet means there is insufficient time to significantly vary the sampling pattern.  The SLM hardware, likely a digital micromirror device (DMD) for broadband capture, can only implement a binary mask.  Discounting the use of a randomly generated mask, on the ground that the associated large sensing matrix can pose unacceptable storage challenges during reconstruction, the constraint of the sensing matrix being binary-valued restricts the range of candidate deterministic constructed sensing bases, though several options exist (e.g. \cite{amini2011} and \cite{lu2012}).  Ruling out binary schemes requiring pattern changes, such as Zhang et al.'s Fourier approach (\cite{zhang2017}), reduces the field further still.  However, Pastuszczak et al.'s binary representation of discrete noiselets (\cite{pastuszczak2016}),	
a family of functions first described by Coifman et al. in \cite{coifman2001}, which have fast $\mathcal{O}(n\log{}n)$ transforms, is ideal for DMD CS imaging.  Apart from being binary, their translation from noiselets' native complex representation is efficient, requiring only $m+1$ binary samples to determine $m$ complex noiselet coefficients, and noiselets are an excellent CS sensing basis for natural scenes, as exemplified by Cand\`es and Romberg's use of them in \cite{candes2007}.  Noiselets' suitability for sampling images stems from them having provably minimal coherence with Haar wavelets (demonstrated by Tuma and Hurley in \cite{tuma2009}), which in turn are known to permit sparse representations of natural scenes.

In conventional SPC applications, as described by Pastuszczak et al., a scene may be sampled by one basis function at a time, shaping the basis to a 2D mask covering the entire scene, and quickly iterating through a sufficient number of functions to obtain enough samples to achieve reconstruction.  While Noblet et al. note the possibility of adapting the pushframe mask mid-capture for the detection of specific targets, in general iterative patterning is not possible in the orbital pushframe context, because of time constraints.  Instead, at a basic level, one can apply a function to a column of the scene, and by having the adjacent mask column represent a different function, apply this other function to the same scene column at the next time step, and so on across the mask as the scene moves before the camera.  The mask's width is then dictated by the number of basis functions one wishes to use when collecting samples.  The disadvantage of this approach is that one is compressing, and later reconstructing, each scene-column independently: there is no spatial, 2D, image context, and this lack of constraint hinders good reconstruction.
		
One way of exploiting the correlations between adjacent columns is Ouyang et al.'s compressive line sensing (CLS), described in \cite{ouyang2014}.  CLS features a scanning imager \emph{projecting} a cross-track line of varying structured illumination, whose return is captured by a single photomultiplier tube; while such a system is not directly applicable to an Earth observation application, the captured data are comparable to those obtained via SLM masking of an externally illuminated scene.  Their sensing matrix models 2D blocks of recorded lines as featuring a common component with per-line unique components, and applies random per-line masks.  Later work from the same group (\cite{ouyang2017}) augments their previous approach to 2D reconstruction by integrating a Bayesian inference concept.

Block compressed sensing (BCS) by Gan \cite{gan2007}, is another departure from the `whole scene' SPC approach, and conceptually more like the conventional block approach used in JPEG etc.  In BCS a (small) sampling basis is fitted to a square 2D sampling window, 32\,$\times$\,32 pixels being suggested, which is then tiled over the scene.  The tiled windows are independently sampled with a variety of patterns, allowing each tile to be stored and reconstructed separately, while capturing some 2D spatial structure.  Subsequent global optimization reduces blocking artefacts, with this stage being a major focus of later BCS works, such as \cite{mun2009} and \cite{chien2017}.	
Since the pushframe architecture's minimum sampling unit is the sum of a single column, small square sampling blocks are not feasible, but it is possible to construct a column-height window which is several columns wide.  Summing the integrated coefficients of each window's columns would broadly reproduce the BCS technique, but is unnecessarily wasteful of the available spatial information.  Merging columnar sampling with aspects of BCS would allow 2D structure to be captured, and present as a constraint during reconstruction, giving superior results.

We believe the combination of the scanning pushframe device with a block-based compressive sampling strategy to be a unique contribution.  Other `SPC array' works do not feature important aspects of our approach.  Arnob et al.'s system (\cite{arnob2018}) is static, and uses the linear detector to have a set of SPCs where each handles a different spectral range, a diffraction grating having been added just before the detector in an otherwise conventional SPC design, with the whole scene sampled at once.  Fowler's theoretical paper (\cite{fowler2014}) is more relevant, with outlines for compressive pushbroom and whiskbroom devices, but again the aim is hyperspectral capture.  His pushbroom approach requires multiple exposures per step, as a number of masks must be displayed, with the sensors of the array independently capturing the spectrum of one spatial pixel each per pushbroom step.  The whiskbroom variant uses a static mask, with the whole array sampling one spatial pixel in each exposure.  Henriksson's patent (\cite{henriksson2016}) details a hardware architecture equivalent to that of the pushframe device, but regards the avoidance of physical scanning as one of the invention's advantages.  The mask is described as consisting of varying vertical stripes, the rows displayed on the SLM being identical, with no mention of block-based compression.  In \cite{wang2015} Wang et al., and in \cite{li2019} Li et al., use similarly restrained one-dimensional pattern designs.

In the following Sections we shall first describe our adaptation of binarized noiselets to column-based sensing, and then detail the enhancements of our `columnar BCS'.  \autoref{s:method} explains the implementation of these points, while \autoref{s:results} shows results obtained under both approaches, with simulated data quantifying the effects of block-size and compression parameters, and real-world data illustrating observed performance using a prototype pushframe device.  \autoref{s:pan} discusses an extension of our methods to a multi-spectral setting, using pan-sharpening techniques.  Finally, we discuss our results in \autoref{s:conc}.

\section{Method}
\label{s:method}

CS is typically expressed as the recovery of some original signal $\mathbf{x}$, from a set of measurements $\mathbf{y}$, where $\mathbf{y = \Phi{}x}$, and $\mathbf{\Phi}$ is a (known) sensing matrix, of size $m\,\times\,n$, $m < n$ --- this inequality leading to the compression.  There are infinite solutions for $\mathbf{x}$, as the system is underdetermined, but the constraint that $\mathbf{x}$ has a sparse representation in some basis $\mathbf{\Psi}$, i.e. $\mathbf{x = \Psi{}f}$, where $\mathbf{f}$ is a vector of mostly zero or near-zero coefficients, makes the problem tractable.  Since, by substitution, $\mathbf{y = \Phi\Psi{}f}$, then if $\mathbf{\Psi}$ is known, $\mathbf{f}$ can obtained through optimization, in turn allowing the reconstruction of $\mathbf{x}$.  In the imaging case it is common that $\mathbf{x} = \vvec(\mathbf{X})$, where $\vvec(\mathbf{X})$ denotes the vectorization of the 2D image $\mathbf{X}$ (of width $w$ and height $h$) by vertically stacking all of $\mathbf{X}$'s columns into a single vector.

As Pastuszczak et al. explain, fewer measurements, $m$, are required as the mutual coherence of $\mathbf{\Phi}$ and $\mathbf{\Psi}$ decreases; natural scenes compress well in the Haar wavelet domain, making it a good candidate basis for $\mathbf{\Psi}$; and noiselet functions have minimal coherence with the Haar basis.  Noiselet matrices are unitary,	
with orders of the form $2^q$, and a $\mathbf{\Phi}$ matrix may be simply obtained by choosing random rows from an appropriately sized noiselet matrix (assuming $n$ is a power of two
) in proportion to the desired compression ratio.  Therefore the noiselet family of functions are a good CS imaging measurement basis.  But noiselets are complex valued, and can have -1, 0 and +1 magnitudes in both their real and imaginary parts.  In \cite{pastuszczak2016} Pastuszczak et al. detail transformations such that an $m\,\times\,n$ noiselet sensing matrix $\mathbf{\Phi}$ can be replaced by a $(m\,+\,1)\,\times\,n$ matrix $\mathbf{P}$, where the elements of $\mathbf{P}$ are binary valued --- suitable for DMD patterning --- and samples $\mathbf{\tilde{y}}$ may be taken by
\begin{equation}
	\label{eq:ypx}
	\mathbf{\tilde{y} = Px}\text{.}
\end{equation}
A reverse transformation converts $\mathbf{\tilde{y}}$ to complex-valued $\mathbf{y}$, suitable for standard CS reconstruction algorithms, as long as a measurement of $\mathbf{X}$'s mean intensity is available for scaling.  Such a measurement can be achieved by adding an all-ones row to $\mathbf{P}$, hence the $m\underline{+1}$ rows stated above.

\subsection{Pushframe binarized noiselets}
\label{ss:pbn}

Equation (\ref{eq:ypx}) may be decomposed into $\tilde{y}_i = \langle{}\mathbf{p}_i, \mathbf{x}\rangle$, where $\tilde{y}_i$ ($1 \leq i \leq m$) is a single measurement coefficient, and $\mathbf{p}_i$ is the $i$th row of $\mathbf{P}$.  In SPC sensing the whole of the scene $\mathbf{X}$ is sampled by one $\mathbf{p}_i$ at any time, and a single $\tilde{y}_i$ recorded.  That is, $n$, the number of columns in $\mathbf{\Phi}$ and $\mathbf{P}$, is equal to $wh$, the number of elements in $\mathbf{X}$.  The dot product summation is physically achieved by all the light transmitted through the $\mathbf{p}_i$ mask being integrated by a single photodiode.

\begin{figure}
	\centring
	\begin{tikzpicture}
		\colorlet{cloudy}{violet!60}
		\node[cloud, cloud puffs = 11, draw, fill=cloudy, minimum width=5cm, minimum height=3cm] at (2, 2) {};
		\node at (5.681, 2) [draw, single arrow, fill = white, minimum width = 1.83cm]{Imager motion};

		\def\pixelmap{{%
			{0,1,1,1,1,0,0,0,1},
			{1,0,1,1,0,1,0,0,1},
			{1,1,0,1,0,0,1,0,1},
			{1,1,1,0,0,0,0,1,1},
			{1,1,1,0,1,1,1,0,1},
			{1,1,0,1,1,1,0,1,1},
			{1,0,1,1,1,0,1,1,1},
			{0,1,1,1,0,1,1,1,1}}}

		\foreach \x [count=\i from 0] in {0.25, .75, ..., 4.5}
			\foreach \y [count=\j from 0] in {0.25, .75, ..., 4}{
				\pgfmathsetmacro{\bit}{\pixelmap[7-\j][\i]}
				\if\bit1
					\node[rectangle, draw, minimum size = 5mm] at (\x,\y) {};
				\else
					\node[fill, rectangle, draw, minimum size = 5mm] at (\x,\y) {};
				\fi
			}

		\foreach \x [count=\i from 1] in {.25, .75, ..., 4.5} {
			\node at (\x, 4.3) {\footnotesize{$\mathbf{p}_\i$}};
			\node at (\x, 1.55) [draw, single arrow, shape border rotate = 270, minimum height = 4.5cm, fill = cloudy!50, single arrow head extend = 1mm, inner xsep = .7mm]{};
			\node at (\x, 1.5) {\tiny$\Sigma$};
		}
		\draw [<-, thick] (4.5,-.3) -- (5.5, -.9);
		\node [align = center] at (6.3, -1.4) {Light integrating\\system};

		\node [rectangle, fill = cloudy!67!white!75!black, draw, minimum size = 5mm] at (.25, -1) {\scriptsize$\tilde{y}_1$};
		\node [rectangle, fill = cloudy!65!white!75!black, draw, minimum size = 5mm] at (.75, -1) {\scriptsize$\tilde{y}_2$};
		\node [rectangle, fill = cloudy!50!white!75!black, draw, minimum size = 5mm] at (1.25, -1) {\scriptsize$\tilde{y}_3$};
		\node [rectangle, fill = cloudy!60!white!75!black, draw, minimum size = 5mm] at (1.75, -1) {\scriptsize$\tilde{y}_4$};
		\node [rectangle, fill = cloudy!65!white!50!black, draw, minimum size = 5mm] at (2.25, -1) {\scriptsize$\tilde{y}_5$};
		\node [rectangle, fill = cloudy!60!white!50!black, draw, minimum size = 5mm] at (2.75, -1) {\scriptsize$\tilde{y}_6$};
		\node [rectangle, fill = cloudy!55!white!50!black, draw, minimum size = 5mm] at (3.25, -1) {\scriptsize$\tilde{y}_7$};
		\node [rectangle, fill = cloudy!50!white!50!black, draw, minimum size = 5mm] at (3.75, -1) {\scriptsize$\tilde{y}_8$};
		\node [rectangle, fill = cloudy!30!white, draw, minimum size = 5mm] at (4.25, -1) {\scriptsize$\tilde{y}_9$};

		\node at (2.25, -1.6) {Linear detector};
	\end{tikzpicture}
	\caption{Pushframe capture, applying sensing basis functions as columns in a mask over the scene (purple cloud), where scene intensities through non-blocked column elements are summed into a detector array, giving the coefficients used in reconstruction}
	\label{f:pf}
\end{figure}

In pushframe sampling the summation occurs separately for each column of $\mathbf{X}$, each sum only capturing a part of the effect of the overall mask, and each generating a distinct coefficient.  A simple way of exploiting this flexibility is to generate $\mathbf{\Phi}$ and $\mathbf{P}$ such that $n$ is equal to $h$, similar to BCS having block-sized sensing matrices; in the above equations $\mathbf{x}$ is then no longer a vectorized version of $\mathbf{X}$, but a single column of it.  By displaying all $m + 1$ sensing row vectors \emph{as columns} adjacent to each other on the SLM (pattern `one' values transmit/reflect, pattern `zero' values block), the dot product of each column of $\mathbf{X}$ with all $m + 1$ vectors will be recorded after the capturing device has taken $m + w$ exposures as it steps in a pushbroom-like manner across the scene (see \autoref{f:pf}).  Each column of $\mathbf{X}$ can then be reconstructed independently, once the coefficients are transformed to complex numbers.  In a theoretical realization, such as that used in simulation, the complex coefficients used in the reconstruction of a certain column of $\mathbf{X}$ are simply those obtained by left-multiplying that column by $\mathbf{\Phi}$, i.e.
\begin{equation}
	\label{eq:yphix}
	\mathbf{y}_j = \mathbf{\Phi{}x}_{*,j}
\end{equation}
where $\mathbf{y}_j$ is the vector of complex coefficients needed for the reconstruction of column $j$ of the scene, and $\mathbf{x}_{*,j}$ is the $j$th column vector of $\mathbf{X}$.

In reality, the various binarized noiselet rows $\mathbf{p}_i$ are applied to a given scene column at different timesteps, while during each exposure the rest of $\mathbf{P}$ is masking the adjacent scene columns.  The resulting coefficients may easily be inserted in a $(2m+w)\,\times\,(m+1)$ matrix $\mathbf{S}$, where the first exposure's coefficients form the main diagonal, and the corresponding ones of subsequent exposures are inserted with an incrementing row position, as illustrated for $m = 4$ in \autoref{f:minsert}.  Note that the Figure's arrangement assumes relative motion as shown in \autoref{f:pf} --- should the scene instead move from $\mathbf{p}_1$ towards $\mathbf{p}_m$, $\mathbf{S}$'s columns would need flipping in the left-right direction.  Once a full set of samples has been recorded, the rows with unset elements can be discarded, and, of the remaining rows, the nth row contains the coefficients appropriate to reconstruct the nth column of the scene.

\begin{figure}
	\setlength\arraycolsep{2pt}
	$$\mathbf{S} = \begin{bmatrix}
\tilde{y}_1\rvert_{_{t=1}}\\
\tilde{y}_1\rvert_{_{t=2}}&	\tilde{y}_2\rvert_{_{t=1}}\\
\tilde{y}_1\rvert_{_{t=3}}&	\tilde{y}_2\rvert_{_{t=2}}&	\tilde{y}_3\rvert_{_{t=1}}\\
\tilde{y}_1\rvert_{_{t=4}}&	\tilde{y}_2\rvert_{_{t=3}}&	\tilde{y}_3\rvert_{_{t=2}}&	\tilde{y}_4\rvert_{_{t=1}}\\
\tilde{y}_1\rvert_{_{t=5}}&	\tilde{y}_2\rvert_{_{t=4}}&	\tilde{y}_3\rvert_{_{t=3}}&	\tilde{y}_4\rvert_{_{t=2}}&	\tilde{y}_5\rvert_{_{t=1}}\\
&				&				\vdots\\
\tilde{y}_1\rvert_{_{t=4+w}}&	\tilde{y}_2\rvert_{_{t=3+w}}&	\tilde{y}_3\rvert_{_{t=2+w}}&	\tilde{y}_4\rvert_{_{t=1+w}}&	\tilde{y}_5\rvert_{_{t=w}}\\
&				\tilde{y}_2\rvert_{_{t=4+w}}&	\tilde{y}_3\rvert_{_{t=3+w}}&	\tilde{y}_4\rvert_{_{t=2+w}}&	\tilde{y}_5\rvert_{_{t=1+w}}\\
&				&				\tilde{y}_3\rvert_{_{t=4+w}}&	\tilde{y}_4\rvert_{_{t=3+w}}&	\tilde{y}_5\rvert_{_{t=2+w}}\\
&				&				&				\tilde{y}_4\rvert_{_{t=4+w}}&	\tilde{y}_5\rvert_{_{t=3+w}}\\
&				&				&				&				\tilde{y}_5\rvert_{_{t=4+w}}
	\end{bmatrix}
	\begin{matrix}
		\\
		\\
		\\
		\\
		\rightarrow\mathbf{y}_1\\
		\phantom{\vdots}\\
		\rightarrow\mathbf{y}_w\\
		\\
		\\
		\\
		\relax
	\end{matrix}$$
	\caption{Arrangement of sample coefficients, for $m = 4$}
	\label{f:minsert}
\end{figure}

Having sampled the scene, there are many possible choices of algorithm for CS-based image recovery, from conventional optimization methods, such as Pati et al.'s OMP (\cite{pati1993}), or Needell and Tropp's CoSaMP (\cite{needell2009}), to deep learning schemes, e.g. Higham et al.'s (\cite{higham2018}).  In this work, we prefer to take a total variation (TV) minimization approach,	
biasing the reconstruction process against high variance (i.e. noisy) solutions: as noted in Needell's later paper on TV (\cite{needell2013}), compared to more rigorously justified methods `total variation minimization gives better empirical image reconstruction results'.  We use Becker et al.'s NESTA algorithm (\cite{becker2011}), which combines a TV regularization step with Nesterov's highly efficient minimization algorithm (\cite{nesterov2005}), however other minimization algorithms could equally be employed, following the sampling methods described here.

\subsection{Columnar BCS}
\label{ss:cbcs}

The obvious weakness in the previous subsection's scheme is the sampling and reconstruction of each scene column in isolation: no 2D spatial constraint is applied, despite natural images tending to have strong correlation between one column and the next.  This weakness is not present in the `whole scene at once' sampling approach of an SPC, but such sampling cannot be easily be realized on a continuously moving pushframe sensor.  In this subsection we present an enhanced approach, where the samples from multiple columns are combined to allow reconstruction of larger 2D blocks.

A na\"ive implementation of the above proposal --- forming a $b$ columns-wide block --- simply gathers $b$ times as many samples, from adjacent columns, to pass as a stacked vector to the minimization algorithm, sets the dimensions for the TV minimization step to be $b$ pixels wide, rather than a single column, and supplies an appropriately updated $\mathbf{\Phi}$.  Adopting the multiplicative approach seen in (\ref{eq:yphix}), the new $\mathbf{\Phi}$ must be compatible with $\vvec(\mathbf{y}_{[j,j+b)}) = \mathbf{\Phi}\vvec(\mathbf{x}_{*,[j,j+b)})$, using interval-like notation to replace the previous references to column $j$, with $[j,j+b)$ indicating the $b$-wide block of columns starting at column $j$.  If the \autoref{ss:pbn} $\mathbf{\Phi}$ is now denoted by $\mathbf{\Phi}_1$, the simplest way to construct the new $\mathbf{\Phi}$ is to repeat $\mathbf{\Phi}_1$ $b$ times along the diagonal of a $bm\,\times\,bn$ matrix.  For example, to sample a vectorized four columns-wide block
\begin{equation}
	\mathbf{\Phi} = \begin{bmatrix}
\mathbf{\Phi}_1&	&	&	\\
&	\mathbf{\Phi}_1&	&	\\
&	&	\mathbf{\Phi}_1&	\\
&	&	&	\mathbf{\Phi}_1\\
	\end{bmatrix}\text{.}
\end{equation}
It may be seen that if $\mathbf{\Phi}$ were not sparse, using such a sensing matrix would require the summation of coefficients from all the columns in the block, as in conventional BCS.  By specializing our $\mathbf{\Phi}$, so each sub-matrix is applied to one scene column at a time, we save these additions, and retain more spatial information for reconstruction, resulting in better performance than if we had filled the matrix with $n = bh$ noiselets.


Our more sophisticated proposal however does not repeat $\mathbf{\Phi}_1$.  As described at the start of this Section, $\mathbf{\Phi}_1$ was created by drawing $m$ rows from an $n\,\times\,n$ noiselet matrix $\mathbf{N}$, where $m < n$.  There are therefore other possible length $n$ noiselet rows which are not present in $\mathbf{\Phi}_1$.  As different noiselet rows capture different aspects of a sampled scene column, and, on average, adjacent scene columns have high correlation, sampling scene column $j+1$ with a noiselet row not used in sampling column $j$ may still provide information relevant to the reconstruction of column $j$, as long as the image recovery algorithm does not process the columns independently.  So treating the $n$ available rows of $\mathbf{N}$ as a pool, we form $\mathbf{\Phi}_2$ using rows not used in $\mathbf{\Phi}_1$.  Clearly the number of unused rows depends on the chosen compression ratio, i.e. how much smaller $m$ is than $n$: for small $m$ there may still be unused rows to be drawn when forming $\mathbf{\Phi}_3$, otherwise, at some point we start redrawing from the pool those least recently used, until we have constructed $\mathbf{\Phi}_b$.  These sensing sub-matrices are then placed in our block-diagonal $\mathbf{\Phi}$:
\begin{equation}
	\mathbf{\Phi} = \begin{bmatrix}
\mathbf{\Phi}_1&	&	&	\\
&	\mathbf{\Phi}_2&	&	\\
&	&	\ddots&	\\
&	&	&	\mathbf{\Phi}_b\\
	\end{bmatrix}\text{.}
\end{equation}


It is apparent that $\mathbf{\Phi}$ grows quickly with $b$: if $b$ were 16, $n$ 256, and a compression ratio of 40\% (i.e. 2.5:1) gave an $m$ of 102, the sensing matrix would have over six million elements, albeit with high sparsity.  Fortunately, no such impractically large numbers need apply to a hardware sensing implementation.  We simply display the entire pool of (binarized) noiselet rows, together with an all-ones column, on the SLM as an $n\,\times\,(n+1)$ array, and then map the relevant measured coefficients from each exposure to those rows allocated to each of the $b$ columns in each block.  To achieve image recovery, $b$ rows of the cropped sample matrix $\mathbf{S}$ are passed as a single vector to the minimization algorithm, and the output vector from this is reshaped to a $h\,\times\,b$ image block, and reassembled with adjacent blocks to form the image.

\begin{figure}
	\hfill
	\subfloat[A full $n = 256$ binarized noiselet matrix, following Pastuszczak et al.'s mapping]{\includegraphics[width=.45\columnwidth]{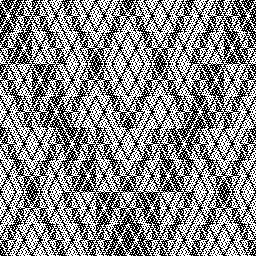}\label{f:Ppast}}
	\hfill
	\subfloat[A full $n = 256$ binarized noiselet matrix, with mirrored mapping]{\includegraphics[width=.45\columnwidth]{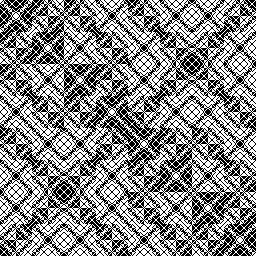}\label{f:Pmirr}}
	\hfill
	\caption{Alternative SLM patterns.  (b) is more optically robust, empirically.}
\end{figure}

\begin{figure*}
	\hfill
	\subfloat[`Blue Jay']{\includegraphics[width=.3\textwidth]{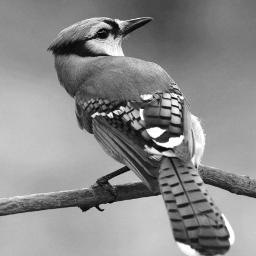}\label{f:bjg}}
	\hfill
	\subfloat[`Gr\'ezac']{\includegraphics[width=.3\textwidth]{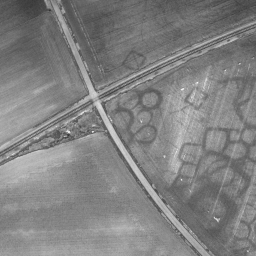}\label{f:grezy}}
	\hfill
	\phantom{.}
	\caption{Test images used in reconstruction simulations}
\end{figure*}

In displaying the entire set of binarized noiselet rows on an SLM, there is free choice for which order the rows appear in --- as described above the coefficients are going to be remapped in any case.  Pastuszczak et al. translate noiselets to binarized noiselets by having the $j$th row of $\mathbf{N}$ map to rows $2j-1$ and $2j$ of the pattern matrix.  Since $\mathbf{N}$ has a canonical row ordering, resulting from the definition of noiselet matrices, a full pattern matrix constructed following Pastuszczak et al.'s mapping can be generated, as shown in \autoref{f:Ppast}.  However, we prefer to use a mirrored mapping, as illustrated in \autoref{f:Pmirr}, where the $j$th row of $\mathbf{N}$ maps to rows $j$ and $n+1-j$ of the pattern matrix.  The main benefit of a matrix formed using this mapping is its superior real-world performance: it seems that the larger contiguous black or white regions mean the degrading effects of optical crosstalk are less severe in experimental data.  A secondary advantage is that Pastuszczak et al.'s binarization scheme requires that where the noiselet row $j$ is used, so is the row $n+1-j$, so our scheme means the same mapping can be employed for tracking the row indices used both with $\mathbf{P}$ for sampling, and with $\mathbf{\Phi}$ for reconstruction.

\section{Results}
\label{s:results}

\subsection{Simulated}

We begin this section by showing simulated, noise-free, results, as simulation allows us to know exactly how the recovered image should appear, and provide quantitative measures of our algorithm's performance in achieving that aim.  Our simulated results feature the capture and reconstruction of the two images below.  The first, `Blue Jay', seen in \autoref{f:bjg}, features a variety of textures, areas of high and low detail, and variations in contrast, making a reasonably challenging subject.  The second, `Gr\'ezac'\footnote{cropped from an aerial photograph by Jacques Dassi\'e, published under the CC BY 2.5 license}, seen in \autoref{f:grezy}, possesses some similar features, but is more representative of Earth-observation satellite imagery, the target application for a pushframe imager.  Both images are 256\,$\times$\,256 pixels in size, this being the highest resolution achieved on an experimental pushframe imager by Noblet et al.  From this it follows that all sensing is performed with $n = 256$.

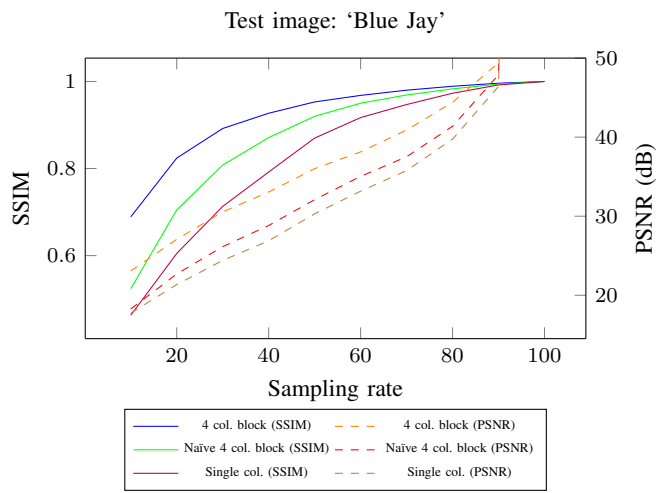
\begin{figure}
	\begin{tikzpicture}
		\begin{axis}[%
			axis y line*=left,
			width=0.93\columnwidth,
			height=0.6\columnwidth,
			hide x axis,
			xmax = 109,
			ylabel=SSIM,
			title={Test image: `Blue Jay'},
			]
			\addplot[color=blue, solid] table [x index = 0, y index = 3, col sep=comma, skip first n = 1] {bj_cr_ssim.csv};
			\label{plt:bjcs1}
			\addplot[color=green, solid] table [x index = 0, y index = 2, col sep=comma, skip first n = 1] {bj_cr_ssim.csv};
			\label{plt:bjcs2}
			\addplot[color=purple, solid] table [x index = 0, y index = 1, col sep=comma, skip first n = 1] {bj_cr_ssim.csv};
			\label{plt:bjcs3}
		\end{axis}
		\begin{axis}[%
			axis y line*=right,
			width=0.93\columnwidth,
			height=0.6\columnwidth,
			xlabel=Sampling rate,
			ylabel=PSNR (dB),
			xmax = 109,
			ymax=50,
			legend style={at={(0.5,-0.25)}, anchor=north},
			legend columns = 2,
			]
			\addlegendimage{/pgfplots/refstyle=plt:bjcs1}\addlegendentry{4 col. block (SSIM)}
			\addplot[color=orange, dashed] table [x index = 0, y index = 3, col sep=comma, skip first n = 1] {bj_cr_snr.csv};
			\addlegendentry{4 col. block (PSNR)}
			\addlegendimage{/pgfplots/refstyle=plt:bjcs2}\addlegendentry{Na\"ive 4 col. block (SSIM)}
			\addplot[color=red, dashed] table [x index = 0, y index = 2, col sep=comma, skip first n = 1] {bj_cr_snr.csv};
			\addlegendentry{Na\"ive 4 col. block (PSNR)}
			\addlegendimage{/pgfplots/refstyle=plt:bjcs3}\addlegendentry{Single col. (SSIM)}
			\addplot[color=brown, dashed] table [x index = 0, y index = 1, col sep=comma, skip first n = 1] {bj_cr_snr.csv};
			\addlegendentry{Single col. (PSNR)}
		\end{axis}
	\end{tikzpicture}
	\caption{Comparison of image recovery performance for the three methods described in \autoref{s:method}, for varying levels of data compression}
	\label{f:methodcomp}
\end{figure}

In \autoref{f:methodcomp}, the comparative performance of three sampling methods from the previous Section can be seen.  The three methods are:
\begin{enumerate}
	\item the basic scheme described in \autoref{ss:pbn};
	\item the `na\"ive' scheme mentioned in \autoref{ss:cbcs}; as we developed towards
	\item our enhanced proposal, using all available noiselet rows, in the same subsection.
\end{enumerate}  For the later two methods a modest block width of four columns is used.  $m$ is varied to change the number of samples taken, which manifests as the level of data compression: 10\% being a reduction to a tenth of the uncompressed image's sample count, and 100\% being no compression.  For each tested configuration the reconstructed image is compared to the uncompressed original, and we measure the peak signal to noise ratio (PSNR) and the more perceptually based structural similarity index measure (SSIM), presented in \cite{wang2004}.  PSNR reaches infinity for an identical image, whereas SSIM takes values between -1 and 1, with 0 indicating no similarity, and 1 perfect similarity.

It is easily seen that by both metrics, the block-based methods outperform the single column approach, and the more advanced block sensing algorithm beats the more basic one, for all tested sampling rates.  \autoref{f:methodcomp} only shows data from the `Blue Jay' image, but the plots from the `Gr\'ezac' image are very similar.

\begin{figure*}
	\hfill
	\begin{tikzpicture}
		\begin{semilogxaxis}[%
			axis y line*=left,
			width=0.4\textwidth,
			height=0.3\textwidth,
			hide x axis,
			ylabel=SSIM,
			title={Test image: `Blue Jay'},
			]
			\addplot[color=blue, solid] table [x index = 0, y index = 2, col sep=comma, skip first n = 1] {bj_ssim.csv};
			\label{plt:bjs1}
			\addplot[color=green, solid] table [x index = 0, y index = 1, col sep=comma, skip first n = 1] {bj_ssim.csv};
			\label{plt:bjs2}
			\addplot[blue, mark=x] plot coordinates { (256, 0.950) };
			\addplot[green, mark=x] plot coordinates { (256, 0.885) };
		\end{semilogxaxis}
		\begin{semilogxaxis}[%
			log ticks with fixed point,
			axis y line*=right,
			width=0.4\textwidth,
			height=0.3\textwidth,
			xtick={1, 2, 4, 8, 16, 32, 64, 128, 256},
			xlabel=2D block width (px),
			ylabel=PSNR (dB),
			legend to name = legned,
			legend columns = -1,
			]
			\addlegendimage{/pgfplots/refstyle=plt:bjs1}\addlegendentry{40\% sampling (SSIM)}
			\addplot[color=orange, dashed] table [x index = 0, y index = 2, col sep=comma, skip first n = 1] {bj_snr.csv};
			\addlegendentry{40\% sampling (PSNR)}
			\addlegendimage{/pgfplots/refstyle=plt:bjs2}\addlegendentry{20\% sampling (SSIM)}
			\addplot[color=red, dashed] table [x index = 0, y index = 1, col sep=comma, skip first n = 1] {bj_snr.csv};
			\addlegendentry{20\% sampling (PSNR)}
			\addplot[orange, mark=x] plot coordinates { (256, 36.466) };
			\addplot[red, mark=x] plot coordinates { (256, 30.361) };
		\end{semilogxaxis}
	\end{tikzpicture}
	\hfill
	\begin{tikzpicture}
		\begin{semilogxaxis}[%
			axis y line*=left,
			width=0.4\textwidth,
			height=0.3\textwidth,
			hide x axis,
			ylabel=SSIM,
			title={Test image: `Gr\'ezac'},
			]
			\addplot[color=blue, solid] table [x index = 0, y index = 2, col sep=comma, skip first n = 1] {grez_ssim.csv};
			\label{plt:gs1}
			\addplot[color=green, solid] table [x index = 0, y index = 1, col sep=comma, skip first n = 1] {grez_ssim.csv};
			\label{plt:gs2}
			\addplot[blue, mark=x] plot coordinates { (256, 0.853) };
			\addplot[green, mark=x] plot coordinates { (256, 0.753) };
		\end{semilogxaxis}
		\begin{semilogxaxis}[%
			log ticks with fixed point,
			axis y line*=right,
			width=0.4\textwidth,
			height=0.3\textwidth,
			xtick={1, 2, 4, 8, 16, 32, 64, 128, 256},
			xlabel=2D block width (px),
			ylabel=PSNR (dB),
			]
			\addplot[color=orange, dashed] table [x index = 0, y index = 2, col sep=comma, skip first n = 1] {grez_snr.csv};
			\addplot[color=red, dashed] table [x index = 0, y index = 1, col sep=comma, skip first n = 1] {grez_snr.csv};
			\addplot[orange, mark=x] plot coordinates { (256, 33.151) };
			\addplot[red, mark=x] plot coordinates { (256, 29.988) };
		\end{semilogxaxis}
	\end{tikzpicture}
	\hfill
	\phantom{.}

	\centring
	\pgfplotslegendfromname{legned}
	\caption{The effect of different 2D block widths on image recovery performance.  The cross ($\times$) marks at the right of the plots are the comparable performance obtained when simulating an SPC-like whole-frame sensing approach.}
	\label{f:blockwidths}
\end{figure*}
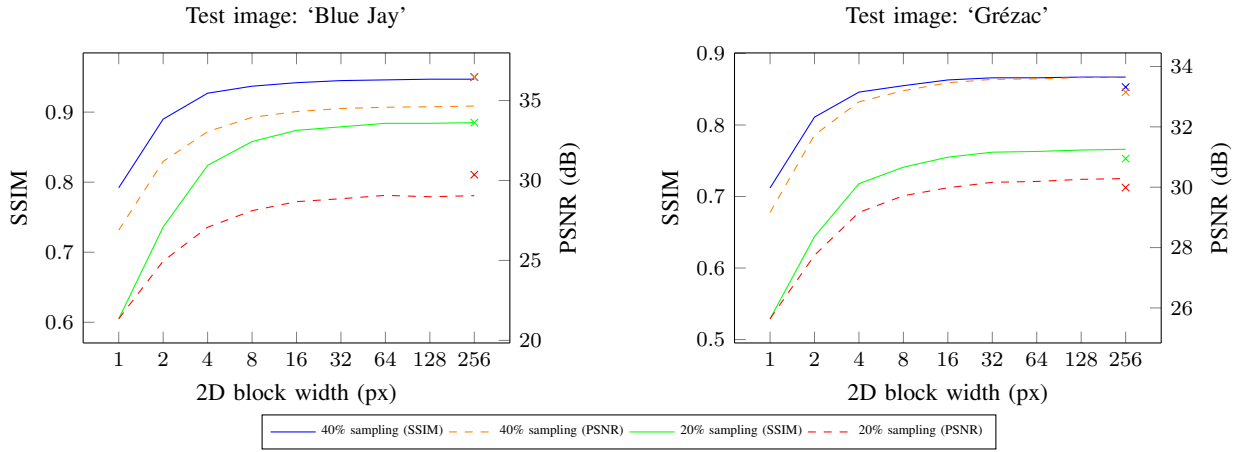

Our columnar block sensing method having been shown to perform well, in \autoref{f:blockwidths} we can see the effect of varying the block width, using the same metrics as before.  The clear finding is that wider blocks perform better.  Intuitively this makes sense: the more blocks that are independently reconstructed, the more errors are likely to occur at the boundaries between blocks, as the optimization process has no visibility over a block's border (hence the research interest in reducing BCS blocking artefacts).  Nonetheless, beyond a certain block size the increase in returns diminishes; the two test images behave a little differently, but the improvement beyond a block width of 16 or 32 pixels is slight.

\autoref{f:blockwidths} shows a maximal block width of 256 pixels --- the size of the sampled image.  For comparison, at the $x = 256$ point of the graphs, crosses have been plotted to show the comparable (binarized noiselet) performance of doing a whole-frame sampling reconstruction (i.e. $n = 65536$), as one might use with a conventional SPC, using the same sampling rates as tested for the pushframe imager.  For the `Blue Jay' image the pushframe block sensing achieves near identical SSIM performance to the whole-frame approach, whereas the whole-frame approach consistently outperforms the other on PSNR by around \SI{2}{dB}.  The situation differs for the `Gr\'ezac' image however, with the pushframe narrowly beating the conventional SPC approach on all metrics.  While finding that the pushframe has a natural affinity to Earth-observation images would be pleasing, it is more likely that the better performer of pushframe and conventional SPC depends highly on the details of any particular scene.

\subsection{Real-world}

\begin{figure}
	\subfloat[Pushbroom-like reference capture]{\includegraphics[width=.24\textwidth]{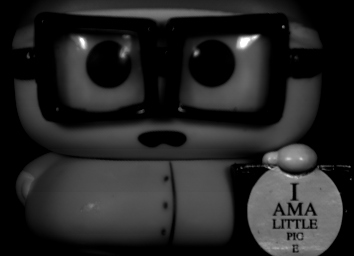}\label{f:cartref}}
	\hfill
	\subfloat[Median-filtered Hadamard matrix capture]{\includegraphics[width=.24\textwidth]{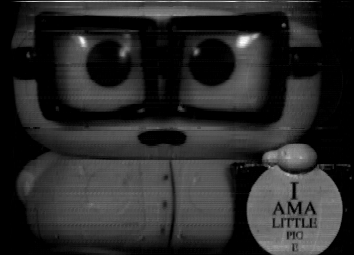}\label{f:cartH}}

	\subfloat[Columnar BCS, 100\% sampled with Pastuszczak's $\mathbf{P}$]{\includegraphics[width=.24\textwidth, angle=180]{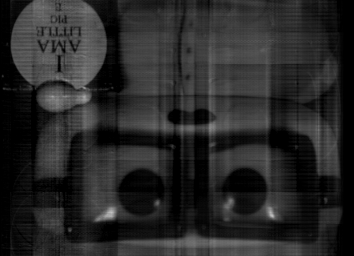}\label{f:cart100nm}}
	\hfill
	\subfloat[Columnar BCS, 100\% sampled with `mirrored' $\mathbf{P}$]{\includegraphics[width=.24\textwidth, angle=180]{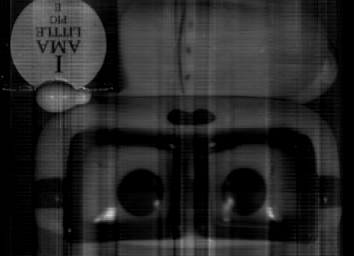}\label{f:cart100}}
	\caption{Comparison of fully sampled scene images}
\end{figure}

For our experimental data, we mount our pushframe imager to a translation stage, causing the scene, approximately \SI{2.5}{m} away, to move sideways from the imager's viewpoint.  We use the same arrangement as in Noblet et al.'s experiments, emulating the optical integration stage by summing the columns captured by a 2D imaging sensor.  This eases experimental alignment and debugging, allows capture of comparison data, and avoids the cost of manufacturing the unusual optics required.  A reference image of the capture subject (a doll of \textit{South Park}'s Eric Cartman character), recorded with the same optics and camera module as used to record the pushframe data, but using the pushbroom-emulation mask mentioned in the introduction, is shown in \autoref{f:cartref}.  This reference is \emph{not} expected to match the recovered images exactly, as the different masks will sample a given part of the scene with the imager at different positions relative to the imaged subject, so perspective and lighting differences are certain.  Our entire optical system uses commercial off-the-shelf components: a Rubinar \SI{300}{mm} f/4.5 telescope, an Epson L3C07U-85G13 LCD module as the SLM, and a FLIR Blackfly S BFS-U3-200S6M camera with a Kowa LM50JC10M \SI{50}{mm} lens mounted via a \SI{22}{mm} extension tube.

As in Noblet et al.'s paper, we derive a partial flat field correction by recording the pushframe coefficients obtained when imaging an all-white scene.  These values are then input as a normalizing calibration curve, weighting subsequently measured coefficients, to compensate for non-uniform optical transmission (vignetting etc.) resulting from the telescope.  This spatial intensity variation is naturally a 2D phenomenon, while the pushframe architecture constrains our correction to only be 1D, so pushframe reconstructions will always be limited by imperfect optics.	

To give a comparison of the sampling and recovery techniques \emph{without data compression}, \autoref{f:cartH}--\ref{f:cart100} feature a previously described technique, and two variants of the enhanced approach from \autoref{ss:cbcs}.  \autoref{f:cartH} shows a reconstruction of the scene using the Walsh-Hadamard sensing matrix approach previously described in Noblet et al.'s paper, using a $3\,\times\,1$ (i.e. purely vertical) median filter to remove the characteristic impulsive rows, rather than the paper's `scrambling' technique.  \autoref{f:cart100nm} and \autoref{f:cart100} were captured using the \autoref{f:Ppast} and \autoref{f:Pmirr} sensing patterns respectively, with $b = n = 256$.  All three recovered images have good qualitative similarity to the reference image, but, of the two CS variants, the superiority of the `mirrored' one is clear, with less horizontal banding in the reconstruction, while the Walsh-Hadamard image suffers some `ghost' artefacts, such as the outline of the character's eyes being present in its chest.  \autoref{t:comp100} gives PSNR and SSIM values for the reconstructions relative to the reference image, but, as noted above, the comparison is not expected to be perfect.

\begin{table}
	\caption{Performance of fully sampled image recovery, relative to reference image}
	\centring
	\label{t:comp100}
	\begin{tabular}{lcc}
		Variant&			PSNR (dB)&	SSIM\\
		\midrule
		Walsh-Hadamard&			26.3&		0.826\\
		CS, Pastuszczak's ordering&	22.1&		0.550\\
		CS, mirrored ordering&		22.9&		0.644
	\end{tabular}
\end{table}

\begin{figure}
	\subfloat[80\% sampling rate]{\includegraphics[width=.24\textwidth, angle=180]{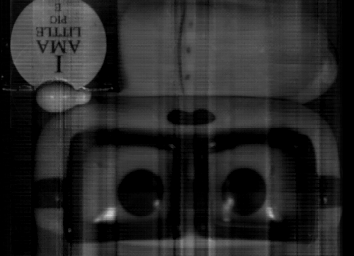}\label{f:cart80}}
	\hfill
	\subfloat[60\% sampling rate]{\includegraphics[width=.24\textwidth, angle=180]{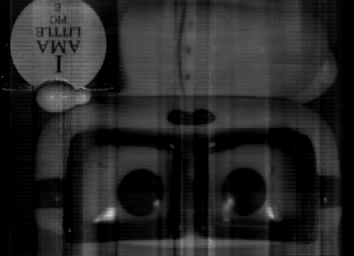}\label{f:cart60}}

	\subfloat[40\% sampling rate]{\includegraphics[width=.24\textwidth, angle=180]{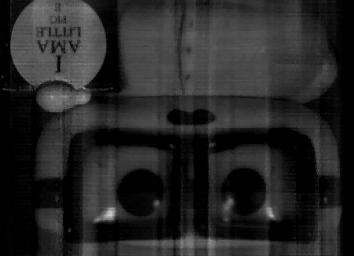}\label{f:cart40}}
	\hfill
	\subfloat[20\% sampling rate]{\includegraphics[width=.24\textwidth, angle=180]{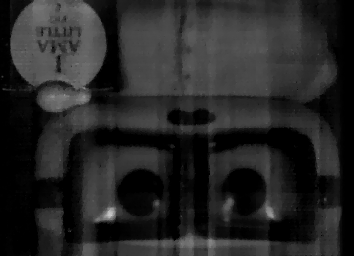}\label{f:cart20}}
	\caption{Effects of reducing sampling rate on columnar-BCS reconstructions}
	\label{f:cartCS}
\end{figure}

Though the filtered Walsh-Hadamard image has the best qualitative performance tabulated, the reconstruction method does not accommodate less than 100\% sampling.  With all other experimental parameters unchanged, \autoref{f:cartCS} shows the effects of reducing the sampling rate, when a mirrored-ordering sensing pattern is used with the \autoref{ss:cbcs} algorithm.  There is little degradation apparent until the sampling rate is below 60\%, even at 40\% the smallest text in the image is clearly legible, and at 20\% the qualitative recovery is still good.  


\section{Multi-spectral extension}
\label{s:pan}

The pushframe architecture elaborated by Noblet et al. can simultaneously capture \emph{intrinsically co-registered} images in multiple wavelengths, by the use of beamsplitting, diffraction of the integrated light on to an area sensor, or a colour (Bayer filtered) camera.  The sampling technique described in \autoref{s:method} can obviously treat wavelengths separately, and reconstruct them independently, but then the data storage and transmission requirement will scale linearly with the number of colour channels.  Images captured at different wavelengths will usually exhibit strong spectral correlation, which is indicative that independent sampling is suboptimal, and some sort of inter-band sensor fusion would yield improved results.

Pushframe multi-spectral data automatically have the same sensing matrix applied in all channels, the incoming scene having interacted with the single SLM before reaching the sensors.  Using the refined block-based sensing approach of \autoref{ss:cbcs}, where the SLM is patterned with all possible binarized noiselet rows, each scene column is in fact sampled with every sensing pattern, and compression ratio is a matter of which samples are discarded.  For multi-spectral samples, the question then is can this complete availability of samples be exploited, by intelligently choosing which coefficients from each wavelength are to be discarded or combined with those from other wavelengths, in order to optimize image recovery following storage or onward transmission of the reduced data-set.  It is possible to retain different coefficients for different bands, capturing a richer variety of spatial data for a given scene column, and then either fuse all the information in a multi-spectral CS reconstruction algorithm (for instance \cite{nagesh2009}, \cite{majumdar2010} or \cite{sugimura2016}),	
or use a multi-spectral super-resolution technique (such as \cite{zomet2002}) on independently reconstructed channels.  However, to more directly demonstrate the possibilities resulting from our fully patterned SLM, we instead show a simple way of achieving superior multi-spectral compression below, inspired by the existing satellite-sensing practice of pan-sharpening.  This is typically used where satellites have a high-resolution broadband `panchromatic' camera, and a number of narrowband cameras of lower resolution, and it is desired to enhance the apparent resolution of the latter.

\begin{figure}
	\subfloat[Colour]{\includegraphics[width=.24\textwidth]{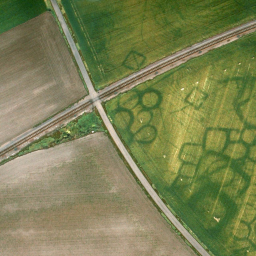}\label{f:grezaccolour}}
	\hfill
	\subfloat[Red]{\includegraphics[width=.24\textwidth]{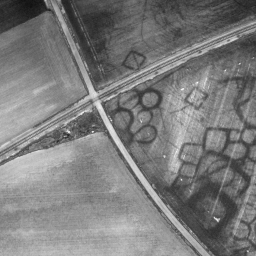}\label{f:grezr}}

	\subfloat[Green]{\includegraphics[width=.24\textwidth]{Grezac256_G}\label{f:grezg}}
	\hfill
	\subfloat[Blue]{\includegraphics[width=.24\textwidth]{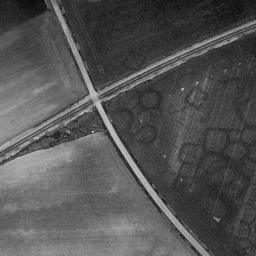}\label{f:grezb}}
	\caption{Colour Gr\'ezac image, along with its RGB decomposition}
\end{figure}

Rather than have a high resolution panchromatic band separately captured from lower resolution colour bands, we synthesize a panchromatic band from the colour band samples before they are compressed further.  More formally, if sample sets $\mathbf{S}_R$, $\mathbf{S}_G$, and $\mathbf{S}_B$ have been collected with $m = n + 1$ in red, green and blue bands respectively, $\mathbf{S}_{pan} = \frac{1}{3}(\mathbf{S}_R+\mathbf{S}_G+\mathbf{S}_B)$, and subsequently only $m_{pan}$ columns of $\mathbf{S}_{pan}$, and $m_{band}$ columns ($m_{band} \ll m_{pan}$) of $\mathbf{S}_R$, $\mathbf{S}_G$ and $\mathbf{S}_B$, are retained.  Standard pan-sharpening techniques (such as those in \cite{vivone2015}) can then be used for recovering high resolution colour bands.  In the following results we use an efficient IHS method, as derived in \cite{tu2001}, where the high frequency differences between the reconstructed panchromatic image and a panchromatic image formed from the reconstructed colour bands are applied straightforwardly to each of the reconstructed colour bands, i.e. $\mathbf{\Delta = I}_{pan} - \frac{1}{3}(\mathbf{I}_R+\mathbf{I}_G+\mathbf{I}_B)$, $\mathbf{I}_R' = \mathbf{I}_R + \mathbf{\Delta}$, and similarly for $\mathbf{I}_G'$ and $\mathbf{I}_B'$.  $\mathbf{I}_{pan}$, $\mathbf{I}_R$, $\mathbf{I}_G$ and $\mathbf{I}_B$ are the recovered panchromatic and band images respectively, while the $\mathbf{I}'$ images are the high resolution versions.  For our simulated capture we use a colour version of the `Gr\'ezac' image (\autoref{f:grezaccolour}), whose similarities and differences between the red, green and blue channels can be seen in \autoref{f:grezr}--\ref{f:grezb}.

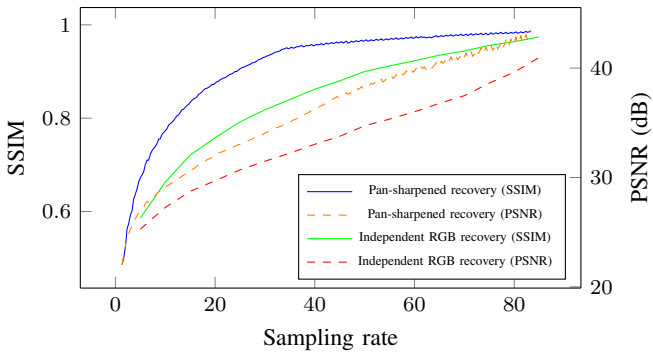
\begin{figure}
	\begin{tikzpicture}
		\begin{axis}[%
			axis y line*=left,
			width=0.93\columnwidth,
			height=0.6\columnwidth,
			hide x axis,
			ylabel=SSIM,
			]
			\addplot[color=blue, solid] table [x expr = \thisrowno{0}/3, y index = 2] {pan-Grezac-sorted_by_ssim};
			\label{plt:gpans}
			\addplot[color=green, solid] table [x index = 0, y index = 2] {grezac_colour_256blk_no-pan};
			\label{plt:gnopans}
		\end{axis}
		\begin{axis}[%
			axis y line*=right,
			width=0.93\columnwidth,
			height=0.6\columnwidth,
			xlabel=Sampling rate,
			ylabel=PSNR (dB),
			legend pos = south east,
			]
			\addlegendimage{/pgfplots/refstyle=plt:gpans}\addlegendentry{Pan-sharpened recovery (SSIM)}
			\addplot[color=orange, dashed] table [x expr = \thisrowno{0}/3, y index = 1] {pan-Grezac-sorted_by_psnr};
			\addlegendentry{Pan-sharpened recovery (PSNR)}
			\addlegendimage{/pgfplots/refstyle=plt:gnopans}\addlegendentry{Independent RGB recovery (SSIM)}
			\addplot[color=red, dashed] table [x index = 0, y index = 1] {grezac_colour_256blk_no-pan};
			\addlegendentry{Independent RGB recovery (PSNR)}
		\end{axis}
	\end{tikzpicture}
	\caption{Comparison of image recovery performance using heterogeneous (overall sampling ratio $\bar{m}$) and homogeneous ($m$) band sampling, using pan-sharpening and independent RGB reconstruction respectively}
	\label{f:panCR}
\end{figure}

The ability to vary $m_{pan}$ and $m_{band}$ independently gives a 2D parameter space for sampling rate, but these may be combined into a single effective value $\bar{m}$, comparable to having a single $m$ applied to independent band sampling/reconstruction: $\bar{m}~=~\frac{m_{pan}}{3}~+~m_{band}$.  In \autoref{f:panCR} the PSNR and SSIM are plotted for the $\bar{m}$ having the highest PSNR or SSIM respectively for a variety of $m_{pan}$ and $m_{band}$ combining to give that $\bar{m}$, along with the PSNR and SSIM resulting from varying $m$ when the bands are independently recovered.

\begin{figure}
	\begin{tikzpicture}
		\begin{axis}[%
			width=0.99\columnwidth,
			height=0.6\columnwidth,
			ylabel style={align = center},
			ylabel=Fraction of samples needed\\with pan-sharpening,
			xlabel style={align = center},
			xlabel={Independent RGB recovery\\sampling rate},
			legend pos = north west,
			]
			\addplot[color=blue, solid] table [x index = 0, y expr = \thisrowno{2}/3/\thisrowno{0}] {pan_cr_comp};
			\addlegendentry{SSIM}
			\addplot[color=orange, dashed] table [x index = 0, y expr = \thisrowno{1}/3/\thisrowno{0}] {pan_cr_comp};
			\addlegendentry{PSNR}
		\end{axis}
	\end{tikzpicture}
	\caption{Illustration of sample count savings made possible by using non-homogeneous band sampling}
	\label{f:panfrac}
\end{figure}
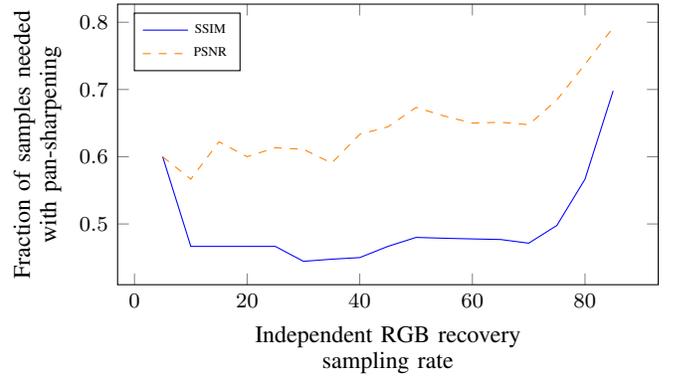

For all tested sampling rates the pan-sharpened approach achieves substantial improvements in both metrics, demonstrating the usefulness of a block-compressing pushframe imager's capability of choosing which samples to retain.  The improvement is further illustrated in \autoref{f:panfrac}, where the proportion of pan-sharpening samples needed to achieve the PSNR/SSIM obtained with independent channel reconstruction is plotted.  For the majority of compression ratios a half to two thirds of the data are required.

\section{Conclusion}
\label{s:conc}

We have described a way of achieving compressive sampling on a pushframe imager.  By using a block-based reconstruction approach, with column-high blocks, our technique delivers good performance, both in simulation and on real-world data, exploiting 2D spatial structure in an imaged scene.  At the same time, it is capable of efficient hardware implementation, through our observation that it is straightforward to sample with a complete binarized noiselet matrix, and then only retain desired combinations of coefficients for image recovery.  Our method of arranging the sampling basis functions obtains superior results to an alternative arrangement on optically captured data.

Our simulations demonstrate that drawing rows from the noiselet basis matrix \emph{uniformly} when sampling each block gives better results than repeated use of a subset of rows, or not using blocks at all.  Furthermore, performance increases as blocks become wider, but gains become more and more marginal.  For offline image recovery, reconstructing very large blocks to gain the best recovery, or using optimization techniques from the BCS literature to remove blocking artefacts, may not be problematic, but in an online situation intermediate block-widths might be more suitable to allow a reconstructed image, still of high quality, to be built up over time.  In a comparison of pushframe and conventional SPC sensing of different scenes there was no clear winner, performance being scene-dependent.  It is fair however from these results to conclude that pushframe block sensing is not necessarily worse than whole-frame SPC sensing, and can be better.

Our experiments showed little deterioration in recovery performance at low compression ratios, though the reconstructions were consistently imperfect.  This performance limit at higher sampling rates is likely to arise from optical nonidealities --- the quality of recovery is very sensitive to the 1D flat field correction curve for instance.  This is also the case for other pushframe reconstruction techniques however, so we can conclude that compressive sampling is especially beneficial when using a pushframe imager: higher data rates confer no advantage in terms of recovered image quality, so to not compress would be wasteful.  Our results show that good recovery is still achieved at lower sampling rates, with the image quality decreasing gracefully as compression increases.

Finally, we have outlined a method for getting even greater data compression when multi-spectral samples are captured.  In itself, the pan-sharpening recovery technique is not sophisticated, but the $\sim$40\% data reduction nonetheless attained is a powerful demonstration of the utility of our enabling pushframe CS approach --- without changing the common mask pattern, variation of different bands' compression levels is very computationally cheap.

\section*{Acknowledgment}

The authors would like to thank Wideblue Limited, Prof. M. Macdonald and Dr. S. R. Owens for useful discussions.

\bibliographystyle{IEEEtran}
\bibliography{paper2}

\begin{thebibliography}{10}
\providecommand{\url}[1]{#1}
\csname url@samestyle\endcsname
\providecommand{\newblock}{\relax}
\providecommand{\bibinfo}[2]{#2}
\providecommand{\BIBentrySTDinterwordspacing}{\spaceskip=0pt\relax}
\providecommand{\BIBentryALTinterwordstretchfactor}{4}
\providecommand{\BIBentryALTinterwordspacing}{\spaceskip=\fontdimen2\font plus
\BIBentryALTinterwordstretchfactor\fontdimen3\font minus
  \fontdimen4\font\relax}
\providecommand{\BIBforeignlanguage}[2]{{%
\expandafter\ifx\csname l@#1\endcsname\relax
\typeout{** WARNING: IEEEtran.bst: No hyphenation pattern has been}%
\typeout{** loaded for the language `#1'. Using the pattern for}%
\typeout{** the default language instead.}%
\else
\language=\csname l@#1\endcsname
\fi
#2}}
\providecommand{\BIBdecl}{\relax}
\BIBdecl

\bibitem{noblet2020}
Y.~Noblet, S.~Bennett, P.~F. Griffin, P.~Murray, S.~Marshall, W.~Roga,
  J.~Jeffers, and D.~Oi, ``Compact multispectral pushframe camera for
  nanosatellites,'' \emph{Applied Optics}, vol.~59, no.~27, pp. 8511--8518,
  Sep. 2020.

\bibitem{candes2006}
E.~J. Cand\`es, J.~K. Romberg, and T.~Tao, ``Stable signal recovery from
  incomplete and inaccurate measurements,'' \emph{Communications on Pure and
  Applied Mathematics}, vol.~59, no.~8, pp. 1207--1223, 2006.

\bibitem{donoho2006}
D.~L. Donoho, ``Compressed sensing,'' \emph{IEEE Trans. Inf. Theory}, vol.~52,
  no.~4, pp. 1289--1306, 2006.

\bibitem{yuan2020}
X.~Yuan and R.~Haimi-Cohen, ``Image compression based on compressive sensing:
  End-to-end comparison with {JPEG},'' \emph{IEEE Trans. Multimedia}, vol.~22,
  no.~11, pp. 2889--2904, 2020.

\bibitem{duarte2008}
M.~F. Duarte, M.~A. Davenport, D.~Takhar, J.~N. Laska, T.~Sun, K.~F. Kelly, and
  R.~G. Baraniuk, ``Single-pixel imaging via compressive sampling,'' \emph{IEEE
  Signal Process. Mag.}, vol.~25, no.~2, pp. 83--91, Mar. 2008.

\bibitem{amini2011}
A.~Amini and F.~Marvasti, ``Deterministic construction of binary, bipolar, and
  ternary compressed sensing matrices,'' \emph{IEEE Trans. Inf. Theory},
  vol.~57, no.~4, pp. 2360--2370, Apr. 2011.

\bibitem{lu2012}
W.~Lu, K.~Kpalma, and J.~Ronsin, ``Sparse binary matrices of {LDPC} codes for
  compressed sensing,'' in \emph{2012 Data Compression Conference}, ser. DCC
  2012, Snowbird, UT, USA, Apr. 2012, p. 405.

\bibitem{zhang2017}
Z.~Zhang, X.~Wang, G.~Zheng, and J.~Zhong, ``Fast {F}ourier single-pixel
  imaging via binary illumination,'' \emph{Scientific Reports}, vol.~7, no.~1,
  p. 12029, Sep. 2017.

\bibitem{pastuszczak2016}
A.~Pastuszczak, B.~Szczygie{\l}, M.~Miko{\l}ajczyk, and R.~Koty\'{n}ski,
  ``Efficient adaptation of complex-valued noiselet sensing matrices for
  compressed single-pixel imaging,'' \emph{Applied Optics}, vol.~55, no.~19,
  pp. 5141--5148, Jul. 2016.

\bibitem{coifman2001}
R.~Coifman, F.~Geshwind, and Y.~Meyer, ``Noiselets,'' \emph{Applied and
  Computational Harmonic Analysis}, vol.~10, no.~1, pp. 27--44, 2001.

\bibitem{candes2007}
E.~Cand{\`{e}}s and J.~Romberg, ``Sparsity and incoherence in compressive
  sampling,'' \emph{Inverse Problems}, vol.~23, no.~3, pp. 969--985, Apr. 2007.

\bibitem{tuma2009}
T.~Tuma and P.~Hurley, ``On the incoherence of noiselet and {H}aar bases,'' in
  \emph{SAMPling Theory and Applications}, ser. SAMPTA'09, Marseille, France,
  May 2009, pp. 243--246.

\bibitem{ouyang2014}
B.~Ouyang, F.~R. Dalgleish, F.~M. Caimi, T.~E. Giddings, W.~Britton, A.~K.
  Vuorenkoski, and G.~Nootz, ``Compressive line sensing underwater imaging
  system,'' \emph{Optical Engineering}, vol.~53, no.~5, p. 051409, May 2014.

\bibitem{ouyang2017}
B.~Ouyang, W.~W. Hou, F.~M. Caimi, F.~R. Dalgleish, A.~K. Vuorenkoski, and
  C.~Gong, ``Integrating dynamic and distributed compressive sensing techniques
  to enhance image quality of the compressive line sensing system for unmanned
  aerial vehicles application,'' \emph{Journal of Applied Remote Sensing},
  vol.~11, no.~3, p. 032407, Jul. 2017.

\bibitem{gan2007}
L.~Gan, ``Block compressed sensing of natural images,'' in \emph{2007 15th
  International Conference on Digital Signal Processing}, ser. DSP 2007,
  Cardiff, Wales, UK, Jul. 2007, pp. 403--406.

\bibitem{mun2009}
S.~Mun and J.~E. Fowler, ``Block compressed sensing of images using directional
  transforms,'' in \emph{2009 IEEE International Conference on Image
  Processing}, ser. ICIP 2009, Cairo, Egypt, Nov. 2009, pp. 3021--3024.

\bibitem{chien2017}
T.~V. Chien, K.~Q. Dinh, B.~Jeon, and M.~Burger, ``Block compressive sensing of
  image and video with nonlocal {L}agrangian multiplier and patch-based sparse
  representation,'' \emph{Signal Processing: Image Communication}, vol.~54, pp.
  93--106, May 2017.

\bibitem{arnob2018}
M.~M.~P. Arnob, H.~Nguyen, Z.~Han, and W.-C. Shih, ``Compressed sensing
  hyperspectral imaging in the 0.9--{\SI{2.5}{\micro\metre}} shortwave infrared
  wavelength range using a digital micromirror device and {InGaAs} linear array
  detector,'' \emph{Applied Optics}, vol.~57, no.~18, pp. 5019--5024, Jun.
  2018.

\bibitem{fowler2014}
J.~E. Fowler, ``Compressive pushbroom and whiskbroom sensing for hyperspectral
  remote-sensing imaging,'' in \emph{2014 IEEE International Conference on
  Image Processing}, ser. ICIP 2014, Paris, France, Oct. 2014, pp. 684--688.

\bibitem{henriksson2016}
M.~Henriksson, ``An imaging system parallelizing compressive sensing imaging,''
  International Patent WO 2016/028\,200 A1, Feb., 2016.

\bibitem{wang2015}
J.~Wang, M.~Gupta, and A.~C. Sankaranarayanan, ``{LiSens} --- a scalable
  architecture for video compressive sensing,'' in \emph{2015 IEEE
  International Conference on Computational Photography}, ser. ICCP 2015,
  Houston, TX, USA, Apr. 2015, pp. 1--9.

\bibitem{li2019}
Y.-H. Li, X.-D. Wang, and Z.~Wang, ``Compressed sensing imaging system based on
  improved theoretical model and its weighted iterative strategy,''
  \emph{Optics Communications}, vol. 439, pp. 76--84, May 2019.

\bibitem{pati1993}
Y.~C. Pati, R.~Rezaiifar, and P.~S. Krishnaprasad, ``Orthogonal matching
  pursuit: recursive function approximation with applications to wavelet
  decomposition,'' in \emph{Twenty-Seventh Asilomar Conference on Signals,
  Systems and Computers}, vol.~1, Pacific Grove, CA, USA, Nov. 1993, pp.
  40--44.

\bibitem{needell2009}
D.~Needell and J.~A. Tropp, ``{CoSaMP}: Iterative signal recovery from
  incomplete and inaccurate samples,'' \emph{Applied and Computational Harmonic
  Analysis}, vol.~26, no.~3, pp. 301--321, May 2009.

\bibitem{higham2018}
C.~F. Higham, R.~Murray-Smith, M.~J. Padgett, and M.~P. Edgar, ``Deep learning
  for real-time single-pixel video,'' \emph{Scientific Reports}, vol.~8, no.~1,
  p. 2369, Feb. 2018.

\bibitem{needell2013}
D.~Needell and R.~Ward, ``Stable image reconstruction using total variation
  minimization,'' \emph{SIAM Journal on Imaging Sciences}, vol.~6, no.~2, pp.
  1035--1058, 2013.

\bibitem{becker2011}
S.~Becker, J.~Bobin, and E.~J. Cand\`es, ``{NESTA}: A fast and accurate
  first-order method for sparse recovery,'' \emph{SIAM Journal on Imaging
  Sciences}, vol.~4, no.~1, pp. 1--39, 2011.

\bibitem{nesterov2005}
Y.~Nesterov, ``Smooth minimization of non-smooth functions,''
  \emph{Mathematical Programming Series A}, vol. 103, no.~1, pp. 127--152, May
  2005.

\bibitem{wang2004}
Z.~Wang, A.~C. Bovik, H.~R. Sheikh, and E.~P. Simoncelli, ``Image quality
  assessment: from error visibility to structural similarity,'' \emph{IEEE
  Trans. Image Process.}, vol.~13, no.~4, pp. 600--612, Apr. 2004.

\bibitem{nagesh2009}
P.~Nagesh and B.~Li, ``Compressive imaging of color images,'' in \emph{2009
  IEEE International Conference on Acoustics, Speech and Signal Processing},
  ser. ICASSP 2009, Taipei, Taiwan, Apr. 2009, pp. 1261--1264.

\bibitem{majumdar2010}
A.~Majumdar and R.~K. Ward, ``Compressed sensing of color images,''
  \emph{Signal Processing}, vol.~90, no.~12, pp. 3122--3127, Dec. 2010.

\bibitem{sugimura2016}
D.~Sugimura, M.~Tomabechi, T.~Hosaka, and T.~Hamamoto, ``Compressive
  multi-spectral imaging using self-correlations of images based on
  hierarchical joint sparsity models,'' \emph{Machine Vision and Applications},
  vol.~27, no.~4, pp. 499--510, May 2016.

\bibitem{zomet2002}
A.~Zomet and S.~Peleg, ``Multi-sensor super-resolution,'' in \emph{Sixth IEEE
  Workshop on Applications of Computer Vision}, ser. WACV 2002, Orlando, FL,
  USA, Dec. 2002, pp. 27--31.

\bibitem{vivone2015}
G.~Vivone, L.~Alparone, J.~Chanussot, M.~D. Mura, A.~Garzelli, G.~A. Licciardi,
  R.~Restaino, and L.~Wald, ``A critical comparison among pansharpening
  algorithms,'' \emph{IEEE Trans. Geosci. Remote Sens.}, vol.~53, no.~5, pp.
  2565--2586, May 2015.

\bibitem{tu2001}
T.-M. Tu, S.-C. Su, H.-C. Shyu, and P.~S. Huang, ``Efficient
  intensity-hue-saturation-based image fusion with saturation compensation,''
  \emph{Optical Engineering}, vol.~40, no.~5, pp. 720--728, May 2001.

\end{thebibliography}

\end{document}